
\doublespace
\overfullrule=0pt
\hsize 6in
\hoffset 0.5in

\def\si{\sigma}
\def\am{{\alpha}^\mu}
\def\a{\alpha}
\def\an{{\alpha}^\nu}
\def\b{\beta}
\def\d{\delta}
\def\m{\mu}
\def\n{\nu}
\def\no{\noindent}

\def\t{\tilde}
\def\w{\omega}

\def\e{\equiv}

\def\la{\lambda}

\def\bra #1{\left\langle#1\right\vert}
\def\ket #1{\left\vert#1\right\rangle}

\Ref\Jevi{
A. Jevicki, `Developments in $2d$ String Theory', Lectures given at the
 Spring School on String Theory, Trieste, April 1993, Brown HET--918 }

\Ref\Kaku{
M. Kaku, {\it Int. J. Mod. Phys.} {\bf A9} (1994) 3103.}

\Ref\Deme{
K. Demeterfi, A. Jevicki and J. P. Rodrigues, {\it Nucl.Phys.}
{\bf B362} (1991) 173; {\bf B365} (1991) 499; {\it Mod. Phys. Lett.}
{\bf A35} (1991) 3199.}
\Ref\Moore{
G. Moore and R. Plesser, {\it Phys. Rev D} {\bf 46} 1730 (1993).}

\Ref\Zwie{
B. Zwiebach, {\it Nucl.Phys.} {\bf B390} (1993) 33;
M. Kaku, {Phys. Rev. D} {\bf 41} (1990) 3734.}

\Ref\Kakuu{
M. Kaku, {Phys. Rev. D} {\bf 49} (1994) 5364. }

\Ref\Poly{
A. M. Polyakov, {\it Mod. Phys. Lett.} {\bf A6} (1991) 635;
Preprint PUPT (Princeton) -1289 (Lectures given at 1991 Jerusalem
Winter School).}

\Ref\Witt{
E. Witten, {\it Nucl. Phys.} {\bf B373} (1992) 187;
I. Klebanov and A. M. Polyakov, {\it Mod. Phys. Lett} {\bf A6}
(1991) 3273. }

\Ref\Miko{
A. Mikovi\'{c}, `Black Holes and Nonperturbative canonical 2d dilaton
gravity',
Imperial-TP/93-94/16.}

\Ref\Natsu{ M. Natsuume and J. Polchinski \journal Nucl.Phys.& B424 (94) 137 }

\Ref\Uros{
B. Uro\v sevi\'c, {\it Phys. Rev. D15} {\bf 48} (1993) 5827.}

\Ref\Callan{ C. G. Callan, D.Friedan, E. J. Martinec and M. J. Perry \journal
Nucl.Phys.& B262 (85) 593}

\Ref\Fran{
P. Di Francesco and D. Kutasov, {\it Nucl. Phys.} {\bf B375} (1992)
119;
P. Di Francesco and D. Kutasov, {\it Phys. Lett.} {\bf B261} (1991)
385.}

\Ref\Uro{
B. Uro\v sevi\'c, {\it Phys. Rev. D15} {\bf 47} (1993) 5460;
I. Ya. Arefeva, P. B. Medvedev, A. P. Zubarev, {\it Mod. Phys. Lett.} {\bf A8}
(1993) 1469; {\bf A8} (1993) 2167.}

\Ref\Samu{
S. Samuel, {\it Nucl.Phys.} {\bf B296} (1988) 187.}

\Ref\Bel{A. Belopolsky and B. Zwiebach, `Off shell closed string amplitudes:
towards a computation of the tachyon potential', Preprint MIT-CPT-2336, 1994.}

\titlepage
\line{\hfill BROWN-HET-972}
\line{\hfill SNUTP-94-118}

\title{{\bf
PERTURBATIVE CALCULATIONS IN THE EFFECTIVE LAGRANGIAN AND $2D$ CLOSED STRING
FIELD THEORY
     }}

\author{Julian Lee \foot{Supported in part by the Korea
Science and Engineering Foundation through the Center for Theoretical Physics.
\ {\bf E-mail: lee@phyb.snu.ac.kr}}}
\address{Center for Theoretical Physics, Seoul National University, Seoul
151-742, Korea}
\andauthor{Branko Uro\v sevi\'c\foot{Supported in part by
the Department of Energy under
contract DE-FG02-91ER40688 -- Task A.
\ {\bf E-mail: urosevic@het.brown.edu}}
 }
\address{\it Physics Department, Brown University, Providence, RI 02912,
USA}

\abstract
This paper is devoted to the study of closed string field theory in two
dimensions. We compare two different approaches: BRST closed string field
theory
and the string effective Lagrangian. We show that the quadratic action and the
pole structures of tree-level scattering
amplitudes agree. We study merits and drawbacks of various gauge fixing
procedures. In particular, we discuss conformal gauge in the context of the
effective Lagrangian, and Siegel and Lorentz-like gauge in the general BRST
approach. We
discuss the ways in which discrete states
survive a particular gauge fixing both by directly solving the equations of
motion, and by analyzing pole structure of the scattering amplitudes.
\vfill
\endpage
\chapter{\bf Introduction}
In the last couple of years two dimensional strings and string--related models
 have been in focus of interest of the theoretical physics
 community [\Jevi]. There are at least two good reasons for that. First, these
are
 the first exactly soluble string models, and second, one might hope that
 better understanding of these simple models may shed some light on how
 to construct a realistic exactly soluble string theory.

 One of the most important practical problems in that direction is to
 better understand the relationship between the simple $2d$ string field
 models
 and the generic BRST string field theory.  The contrast between
 them is
 perhaps most striking in the closed string case [\Kaku]. On one hand,
 the
 success and simplicity of the collective field theory and the free
 fermion models are well--known [\Deme,\Moore]. On the other hand,
it was proven by Kaku that a consistent BRST string field theory in $2d$, just
like its critical dimension counterpart [\Zwie], requires the introduction of
 an infinite tower of vertices even on the classical level [\Kakuu]. Despite
the
apparent abundance of fields, however, only a handful survive the gauge
 fixing since the physical spectrum consists of a massless tachyon and
 the discrete states (DS)[\Poly,\Witt].

No less interesting is the relation between different $2d$ approaches on the
$S$
 matrix level. In fact, string amplitudes are related to the collective
 field $S$-matrix elements by the multiplicative external
 leg Gamma function factors [\Deme]. Although these factors can be interpreted
as pure phases when the external momenta are real, they
give resonance poles at imaginary values of momenta which signal the existence
of the discrete states.

 As a first step towards establishing the bridge between the BRST field theory
and the collective Lagrangian, it is important to study the low energy
effective
string Lagrangian and its connection with the full BRST formalism.

The aim of this paper is to make a careful comparison between the low energy
effective Lagrangian approach, described
by the Dilaton Gravity (DG) coupled to the massless tachyon field [\Miko], and
the BRST closed string field theory. We compare both the quadratic parts of
the
actions and the scattering amplitudes and show that they agree. In the
process,
we present various ways of gauge fixing for $2d$ strings.

In Sec. 2 we calculate the scattering amplitudes in the effective Lagrangian
using
 the momentum space Feynman diagram formalism.
 Recently, calculations based on wave packet scattering in coordinate
 space were performed in [\Natsu]. In comparison, our calculations make
 use of the standard Feynman rules. In this approach some nontrivial
 features appear in particular gauges. In particular, in order to reproduce the
correct
gravitational scattering we show that the conformal gauge constraints should
be
modified. Our results are shown to agree with the results in ref. [\Natsu].

In Sec. 3 we present the gauge invariant formulation of the BRST closed string
field theory (CSFT) in $2d$. We show that, on the first excited level, the
quadratic part of the CSFT action corresponds to the linearized effective
Lagrangian. In Sec. 4 we
study CSFT in Siegel's gauge, used later in the computations of the string
field
amplitudes (Sec. 6). In Sec. 5 we discuss CSFT in Lorentz-like gauge and show
how, out of abundance of off--shell fields, only discrete states identical to
those found in the first quantized approach [\Poly] survive. We conclude (Sec.
6) by
presenting an alternative way of calculating the scattering amplitudes in the
CSFT framework. Our amplitudes, given as sums over poles, are shown to agree
with the tree-level string results, as well as with the effective Lagrangian
calculations in Sec. 2,
to the order of $\mu^0$. Sec. 7 is reserved for the discussion of the results
and final comments.

\chapter{\bf Pole structure of scattering amplitudes in the effective
Lagrangian
}

In this section we study dynamics of the string effective Lagrangian which is
the $2d$ Dilaton Gravity coupled
to
a massless field, denoted often as tachyon. It is given by [\Miko]:

$$ S= {1 \over 2} \int dt dq \sqrt{-G} e^{-2 \Phi} \left[ a_1 \{ R + 4(\nabla
\Phi )^2 + 16 \} - (\nabla T )^2
+ 4 T^2  -{2 \over 3} a_2 T^3 + O(T^4) \right] \eqn\dil$$

\noindent
which can be derived, for example, by the beta function method[\Callan]. Note
that in \dil~ $a_i$ are arbitrary coefficients. We shall exploit this freedom
later to ensure that the $S$ matrix elements obtained from \dil~ agree with
the
standard tree level string results.

Since the Lagrangian \dil~ is invariant under the general coordinate
transformations, there are two redundant degrees of
freedom corresponding to the reparametrization symmetry. In order to derive
the
Feynman rules for the theory, one needs to fix the gauge. An adequate gauge
choice is, as we shall see in detail below, of utmost importance in $2d$ since
the only gauge degrees of freedom in $2d$ are discrete states. In this section
we, following [\Natsu], fix the conformal gauge: $ G^{\mu \nu} = e^{2\rho}
\eta^{\mu \nu }$. Here, $\eta_{\mu\nu}$ is a flat metric: $ \eta_{00} =
-1,\eta_{11}=1, \eta_{10} = \eta_{01} =0$. Substituting this expression for
$G^{\mu \nu}$ back into \dil~, expanding around the linear
 vacuum $ \Phi = 2q + f $, and redefining the tachyon $ T \equiv e^{2q} S $,
 and introducing the light-cone coordinate $x^\pm \e t \pm q$,
 one obtains the following gauge-fixed Lagrangian:

$$\eqalign{
S_{fix} &= {1 \over 2} \int dt dq \ [ a_1 e^{-4q -2 f} \{ 8 \partial_+
\partial_-\rho - 16 \partial_+f \partial_- f + 16 (\partial_+ -
\partial_- ) f + 16(1 + e^{2 \rho} ) \} \cr
&+ e^{-2 f} \{ 4 (\partial_+
S ) (\partial_- S) + 4 S ( \partial_- - \partial_+ ) S + 4(e^{2 \rho} -1 ) S^2
-{2 \over 3} a_2 e^{2 \rho + 2q} S^3 \} +O(S^4) ] \cr} \eqn\fix $$

\noindent
To read-off propagators and vertices, it is useful to separate the action
\fix~  as $\, S \sim S_2 + S_3$:

$$
\eqalign{
S_2 &= \, {1 \over 2} \int dt dq \big[ e^{-4q} a_1\{ -16 ( \partial_+ f ) (
\partial_-
f ) + 32 \rho^2 - 16 f \partial_+ \partial_- \rho  - 64 f
\rho \} + 4 \partial_+ S \partial_-S \big]  \cr
S_3 &= {1 \over 2} \int dt dq \left[ 8 \rho S^2 - 8 f (\partial_+ S)
(\partial_- S) - 8 f S (\partial_- - \partial_+) S - {2 \over 3} a_2
e^{2q} S^3 \right] \cr
}
$$

\noindent
One can further simplify this expression by making a transformation $f= e^{2q}
\t f, \rho=e^{2q} \t \rho $ in order to get rid of the $e^{-4q}$ factor in
$S_2$:

$$  \eqalign{ S_2 &= {1 \over 2} \int dt dq \big[ \ a_1\{ -16 ( \partial_+
\t f
) ( \partial_- \t f ) + 16 \t f^2 + 32 \t \rho^2 - 16 \t f
\partial_+ \partial_- \t \rho -16 \t f (
\partial_- - \partial_+ ) \t \rho - 48 \t f \t \rho \}
\cr
&\qquad \qquad \qquad + 4 \partial_+ S \partial_-S \big] \cr
S_3 &= {1 \over 2} \int dt dq \ e^{2 q} \left[  8 \t \rho S^2 - 8 \t f
(\partial_+ S)
(\partial_- S) - 8 \t f S (\partial_- - \partial_+) S - {2 \over 3} a_2
 S^3 \right] \, . \cr }$$

\noindent
Next, we diagonalize the action using the ansatz:

$$\eqalign{ \t f &=   \zeta +  \hat O_1 \varphi \cr
\t \rho &=  \hat O_2  \zeta + \varphi }$$

\noindent
where $\hat O_1,\hat O_2$ are some differential operators which we need to
identify. It is easy to see that the condition for vanishing of the cross
terms, upon substituting the ansatz into the action, is given by the following
operator equation:

$$
2 \partial_+ \partial_- \hat O_1 + 2 \hat O_1 +
4  \hat O_2^\dagger -  (\partial_+ \partial_- +
\partial_- - \partial _+ + 3 ) -  \hat O_2^\dagger
( \partial_+ \partial_-  +
\partial_+ - \partial _- + 3 ) \hat O_1 = 0 $$

\noindent
This leaves a lot of freedom in the choice of $\hat O_i$ since there is just
one
equation and two unknowns. We may simply define $\hat O_1 =
0$. In that case we obtain:

$$\t f =  \zeta \, \, \, , \, \, \,
\t \rho = \varphi + {1 \over 4} ( \partial_+\partial_-  +
\partial_+ - \partial _- + 3 )\zeta  $$

\noindent
Substituting these expressions back into the Lagrangian, we get:

$$\eqalign
{ S_2 &= \int[ a_1 \{ \zeta ( - \partial_+^2 \partial_-^2 + \partial_+^2
+\partial_-^2 -1 ) \zeta + 16 \varphi^2 \} + 2 \partial_+ S \partial_- S ] \cr
S_3 &= \int \left[ e^{2q} 4 \varphi S^2 + S^2\partial_+\partial_-\{e^{2q} \zeta
\}
- 4 \partial_+S \partial_-S \zeta e^{2q} -{e^{2q} \over 3} a_2 S^3 \right]
\cr } \eqn\fixed$$

To introduce the Liouville background,  we expand $S,f,\rho$ in \fixed~
around
a background $S_0, f_0, \rho_0$ which is the static solution to the equations
of
motion. To the leading order in $e^{2q}$ they are given by

$$\eqalign{ S_0 &= 2b_1 q + b_2 = b_1(x_+ - x_-) + b_2 + O(e^{2q}) \cr
            f_0 &= O(e^{4q}), \rho_0= O(e^{4q}) \cr } $$

In the last expression, $b_1$, $b_2$ are the some constants of the order $\mu$
(the cosmological constant). A priori, $b_i$, just like $a_i$, are
completely arbitrary. Like $a_i$, their values are determined through the
correspondence of the $S$ matrices (see below).

Splitting $S$ as $S = S_0 + \t S$, the tachyon part of the Lagrangian can be
rewritten as:

$$ S_{tachyon} = {1 \over 2} \int e^{-2 f}\left[ 4(\partial_+ \t S)(\partial_-
\t S ) + 4( e^{2\rho} -1) \t S^2 - 2 a_2 e^{2 \rho + (x_+ - x_-)} S_0 \t S^2
-{2a_2 \over 3} e^{2\rho + x_+ - x_-} \t S^3 \right] \eqn\liou $$

\section{\bf Scattering Amplitudes in the effective Lagrangian}

The action \liou~ can be used to calculate the scattering amplitudes in this
approach. In
particular, we are interested in the pole structure of the $S$ matrix. We use
a
conventional Feynman diagram approach in the Fourier space, instead of
coordinate space method advocated in [\Natsu]. In this approach some nontrivial
features appear in particular gauges.

Before we proceed any further, let us note that a generic interaction term in
\fixed~ has an exponential factor of the form
$e^{nq}\  (n>0)
$
whose Fourier transform
$$f_n (k) \equiv {1 \over 2\pi} \int_{-\infty}^\infty dq e^{ik+nq} $$
 is not well defined. To regularize this integral, we
 replace $e^{nq}$ by $e^{nq} \theta(L-q)$, {\it i.e.} we introduce
 a  space cutoff and take the limit $L \to \infty $ at the end of the
calculation. As we shall see, pole structure is independent of the cut off.

In this section,  scattering amplitudes are normalized as in [\Natsu]. The
reduction formula in that case reads:

$$ \eqalign{ S_{1,\cdots n \to n+1,\cdots,n+m}
&= ({1 \over \sqrt 2 } )^{n+m} (\w_1^2 - k_1^2) \cdots (\w_{n+1}^2 - k_{n+1}^2
)(-1)^m (2\pi)^{2(n+m)} \cr
&\times G(\w_1,k_1;\cdots;\w_{n+m},k_{n+m})
}\eqn\reduc$$

\noindent
where $$G(\w_1,k_1;\cdots;\w_{n+m},k_{n+m}) \equiv \langle \t S(\w_1,k_1),
\cdots
\t S(\w_{n+m},k_{n+m}),\rangle$$
and the Fourier components $\t S$ are defined by $$S(t,q) \equiv \int d\w dk
e^{i
(\w t - kq)} \t S(\w,k).$$
In the above
expression, $(\w_i,k_i)$'s represent the external energy-momenta. As usual, it
is implied that we put the external momenta on-shell, $\w_i^2-k_i^2=0$ after
we
calculate the residues.
Note that
$\w_i > 0$  corresponds to an incoming particle with momentum
$k_i$, while $\w_i < 0$  corresponds to an outgoing particle with momentum
$-k_i$. We assume that the incoming particles move to the right while the
outgoing particles move to the left, so that, for the incoming particles, $k_i
>
0$, $\w_i = k_i$, whereas for the outgoing ones $\w_i = -k_i$.

Our next task is to find the pole structure of the effective Lagrangian $S$
matrix, and to
compare
it with the lowest lying poles of the tree-level string $S$-matrix [\Poly,
\Fran]. The later is given by:
$$ S_{1,\cdots,n \to n+1} = 2\pi i \d(k_{n+1}-\sum_{i=1}^n k_i)
\prod_{i=1}^{n+1} {k_i \Gamma(ik_i) \over \Gamma(-i k_i) } { (-ik_{n+1}-1)!
\over (-ik_{n+1}-n+1)!} \mu^{-ik_{n+1}-n+1}. \eqn\strS $$

\noindent
Notice that in this section the notation differs from the rest of the paper
in that matter momenta $k$ have poles at imaginary values, and are rescaled by
the factor of $1 \over \sqrt 2$.

Since the only non-trivial amplitude corresponds to the scattering $n \to 1$ (
or $1 \to n$, which is just a time
reversal of the above), we concentrate on that case. One can easily see that
the
resonance poles of the
$S$ matrix in the effective Lagrangian approach originate from the anomalous
momentum
conservation at the interaction vertices and the on-shell conditions on
intermediate particles at the internal propagators. The anomalous conservation
is due to the $e^{2q}$ factors in the vertices. If we introduce explicitly the
cosmological constant $\mu$, background potential can be expanded in powers of
$\mu e^{2 q}$ (with possible $\log \mu$'s). Then, for an arbitrary $n+1$ point
Feynman diagram one has the following anomalous conservation law:

$$ \sum_{i=1}^{n+1} k_i = 2(v+m)i \eqn\toano $$

\noindent
where $v$ is the number of the cubic vertices, and $m$ is the order of this
diagram
in the power of $\mu$. For a tree graph, standard topological considerations
give:

$$ v-I =1, \quad 3v = 2I+(n+1) \eqn\euler $$

\noindent
Substituting \euler~ into \toano~, one gets:

$$ \sum_{i=1}^{n+1} k_i = 2(n-1+m)i \eqn\toanot $$

\noindent
On the other hand, since for a $n \to 1$ scattering one has:

$$ k_{n+1} = \sum_{i=1}^n k_i   \, .\eqn\energy$$

\noindent
The Eq. \energy~, together with the Eq. \toanot~, yields:

$$ k_{n+1} = (n-1+m)i \, .$$

The on-shell conditions for the intermediate states will give additional poles
in $k_i$'s.

We have now all the necessary ingredients to apply the formalism described
above. Consider, first,  the two-point function. The full tree amplitude is
given by:

$$ \eqalign{ S_{1 \to 2} &= - 2 \pi \d(k_2 - k_1) k_1 \mu^{-i k_2} {
\Gamma(ik_1) \Gamma(ik_2)
\over \Gamma(-ik_1) \Gamma(-ik_2)} \cr
                      &= -2 \pi \d(k_2 - k_1) k_1 \mu^{-i k_2}{ \Gamma^2(ik_2)
\over
                      \Gamma^2(-ik_2) } \cr } \eqn\stw $$

\noindent
which, if one picks the contribution to the lowest lying pole $k_2 = i$ gives:

$$ S_{1 \to 2} \sim - 2 \pi i \d(k_2-k_1) \mu \left[ {1 \over (ik_2+1)^2} +
{1 \over ik_2 +1} (4 \Gamma'(1) +1-\ln \mu ) \right] $$

\noindent
Let us compare this result with the expression obtained from the effective
Lagrangian.
 The 2-point
function, to the first order in $\mu$, is given by the Feynman diagram in
fig.1,
 where the background terms are treated as a quadratic interaction. By the
reduction
formula \reduc, we have:

$$\eqalign{ S_2^{(1)} &= ({1 \over \sqrt 2} )^2 2! i^3 (-1) 2 \pi \d
(\w_1+\w_2)
(-V(k_1+k_2))    \cr
          &= 2 \pi i a_2 \d(k_2-k_1) \left[ {b_1 \over 2} {e^{2(ik_2+1)L}
\over
            (i k_2 +1)^2
          } - (b_1 L + {b_2 \over 2})  {e^{2(ik_2+1)L} \over ik_2 +1} \right]
\cr
          &\sim 2 \pi i a_2 \d(k_2-k_1) \left[ {b_1 \over 2}{ 1 \over
   (i k_2 +1 )^2 } - { b_2 \over 2} { 1 \over ik_2 +1} \right] \cr
           }, \eqn\etw
$$

\noindent
which agrees with \etw~ provided that:

 $$b_1 a_2 = - 2 \mu, \quad b_2 = b_1(-1-4\Gamma'(1)+ \ln \mu ) \eqn \fvar $$

\noindent
It is important to note that the residue does not depend on the cutoff $L$.

Now we calculate the leading pole of the three-point function. The string
S-matrix is in that case given by:

$$ \eqalign{ S_{1,2 \to 3} &= 2\pi i \delta(k_3-k_2-k_1) \prod_{i=1}^3 k_i {
\Gamma(i k_i) \over \Gamma(-i k_i) } \mu^{-i k_3 -1} \cr &\sim 2 \pi i
\delta (k_3-k_2-k_1) i k_2 (i-k_2) { (-1) \over (i k_3 +1) \Gamma(1) } {
\Gamma(i k_2) \over \Gamma(-i k_2) } { \Gamma(-1-i k_2) \over \Gamma(1 +
ik_2) } \mu^0 \cr &= 2 \pi i \delta (k_3-k_2-k_1) i k_2 (i-k_2) { (-1) \over
(i
k_3
+1) (ik_2) (-ik_2-1) } \cr &= { 2\pi i \over k_3 -i} \delta(k_3-k_2-k_1) \cr }
\eqn\sth$$

\noindent
Again, we keep here just terms of the order $\mu^0$. One can easily convince
oneself that the other poles in $k_3$ correspond to the higher orders in
$\mu$.
We may now compare this expression with the effective Lagrangian 3-point
function to the same order in $\mu$ (fig.2). We have:

 $$ \eqalign{ S_{1,2 \to 3}^{(0)} &= 3!({1 \over \sqrt 2 })^3 i^4 (-1) (2
\pi)^2
\d(k_3 -k_2 -k_1)  f_2 (k_1 + k_2+k_3)  \cr
      &= ({1 \over \sqrt 2})^3 { 4 \pi a_2 e^{(i(k_1+k_2+k_3)+2)L} \over
i(k_1+k_2 +k_3)+2}
       \d(k_3-k_2-k_1)  \cr
       &\sim 2 \pi i \d(k_3-k_2-k_1) {1 \over k_3-i} (-{a_2 \over 2
\sqrt 2}) \cr}
       \eqn\eth $$

\noindent
We see that \sth~ and \eth~ agree if $a_2 = -2\sqrt
2$.

So far, we have seen that $2$ and $3$ point amplitudes agree with
 the corresponding string results. In the process, we have fixed the
 values of some of the, initially arbitrary, coefficients. It is,
 therefore, important to check that our results are consistent for
 the four-point functions as well. That would give the first
 completely non-trivial check of the formalism. To that end, let
 us consider:

  $$ \eqalign{ S_{1,2,3 \to 4} &= 2\pi i \delta(k_4-k_3-k_2-k_1) \prod_{i=1}^4
k_i (-ik_4 -1) {
\Gamma(i k_i) \over \Gamma(-i k_i) } \mu^{-i k_4 -2} \cr
        &\sim 2 \pi i
\delta (k_4- k_3-k_2-k_1) 2i k_1 k_2 (2i-k_1 -k_2) { 1 \over 2! (i k_4 +2)
\Gamma(2) }  \cr
& \times { \Gamma(i k_1)\Gamma(i k_2) \Gamma(-2-ik_1-ik_2) \over
\Gamma(-ik_1) \Gamma(-i k_2) } { \Gamma(2+ik_1 + k_2) \over   \Gamma(1 +
k_2) } \mu^0 \cr
      &= 2 \pi i \delta (k_4-k_3-k_2-k_1)  (2i-k_1-k_2) k_1 k_2  \cr
      &\times {
\Gamma(ik_1) \Gamma(ik_2) \Gamma(-2-ik_1 -ik_2) \over (k_4-2i) \Gamma(-ik_1)
\Gamma(-ik_2)
      \Gamma(2+ik_1+ik_2) } \cr } \eqn\efo $$

\noindent
If we pick the lowest lying pole in $k_1$, we get:

  $$  \eqalign{ S_{1,2,3 \to 4} &\sim 2 \pi i \d (k_4-k_3-k_2-k_1) (i - k_2)
ik_2 {(-1) (i k_2-1)! (-2-ik_2)! \over (k_4-2i) (ik_1+1)\Gamma(1) (-ik_2-1)!
  (ik_2)!} \cr
  &= 2 \pi i \d (k_4 -k_3 -k_2 -k_1) {1 \over (k_4 -2i) (k_1-i) } \cr }
\eqn\sft$$

\noindent
Similarly, we can pick lowest poles in $k_2, k_3=2i-k_1-k_2$ to obtain:
$$ S_{1,2,3 \to 4} \sim 2 \pi i \d (k_4 -k_3 -k_2 -k_1) {1 \over (k_4 -2i)}
\left( {1 \over k_1-i } + {1 \over k_2-i } + {1 \over k_3-i } \right)
\eqn\sfth $$

\noindent
We would now like to compare this result with the Feynman diagram of the
effective Lagrangian.
Consider the diagram due to the tachyon exchange shown in fig.3.

$$ \eqalign{ S_{1,2,3 \to 4}^T &= ({1 \over \sqrt2 })^4 (-1) 3!^2 i^7 (2
\pi)^2
\int dk f_2 (k_1+k_2+k) {1 \over (k_1+k_2)^2 -k^2 +i \epsilon} \cr
      &\times f_2 (k_3+k_4-k) \d
(k_4-k_3-k_2-k_1) \cr
\cr
&= ({1 \over \sqrt 2})^4 i (2a_2)^2 \int {dk \over (i(k_1+k_2 +k)
+2)[(k_1+k_2)^2-k^2 +i\epsilon ]
(i(k_3+k_4-k)+2) } \cr
   & \times \d(k_4-k_3-k_2-k_1) e^{(i(k_1+k_2+k_3+k_4)+4)L} \cr
  }$$

\noindent
where it is assumed that we sum over the inequivalent diagrams.
The poles for $k$ are at $k=k_1+k_2+\epsilon i, k=-(k_1+k_2) + 2 i $. The
integral over $k$ can be calculated by appropriately closing the contour and
picking up the residues in these poles. One obtains:

$$ \eqalign{S_4^T &= 2 \pi a_2^2 \d
(k_4-k_3-k_2-k_1)e^{(i(k_1+k_2+k_3+k_4)+4)L}[ {1 \over 4
(k_1+k_2+\epsilon i) (k_1+k_2-i)} \cr
&\times{1 \over k_1+k_2-k_3-k_4+2i }+{1 \over 2i
(-2(k_1+k_2)+2i)(-k_1-k_2-k_3-k_4 +4i) } ]\cr \cr
   &= -{2 \pi i a_2^2 \over 8 } {k_3-2i \over (k_4-2i)(k_3-i)(k_4-k_3+i
\epsilon) } e^{2(ik_4+2)L} \cr \cr
   &\sim {2 \pi i a_2^2 \over 8} {1 \over (k_4-2i)(k_3-i)} \cr}$$

\noindent
 Substituting the value $a_2 = -2 \sqrt 2$ obtained above and summing over the
inequivalent diagrams, we get

 $$ S_4^T \sim {2 \pi i \over k_4-2i} \d(k_4-k_3-k_2-k_1) \left[ {1 \over
k_3-i}
+
{1 \over k_2-i} +
{1 \over k_1-i} \right]  \eqn\perm $$

\noindent
which agrees with \sfth~. Note that in arriving at \perm~, we have first
picked
a pole in $k_4$, treating $k_3$ as a real
number, and then picked a pole in $k_3$. This coincides with the prescription
used by
Natsuume and Polchinski in ref [\Natsu] in that, there, they first extracted
the
leading contribution from the whole wave packet tail, and only then extracted
the contribution from the separating wave packets.

We could, in principle, consider the contribution to the four-point function
coming from the quartic tachyon vertex (fig.4). To get the correct anomalous
momentum
conservation which gives a pole at $k_4 = 2i$, we easily see that it should be
of
the form $e^{4q} S^4$. However it does not have a propagating intermediate
state
and therefore it does not lead to any additional poles in $k_1,k_2,k_3$.

\section{\bf Scattering With the Exchange of Gravitons}

We next consider scattering with the exchange of gauge `particles'. We put the
word `particles' in quotes since, of course, there is no gauge particles in
two
dimensions. This makes the procedure quite subtle. As we shall see, in order
to
obtain a sensible result, one needs to take a closer look at the conformal
gauge.

It is obvious that the potential contribution from the $\varphi$ intermediate
state
is trivial since it is a just a Lagrange multiplier. Much more interesting is
the exchange of $\zeta$ (fig.5) whose contribution to the lowest order is
given by:

$$\eqalign{ S_\zeta &= ({1 \over \sqrt 2})^4 (-{i \over
a_1})\d(\w_4+\w_3-\w) \d(\w_1+\w_2+\w) e^{(i(k_1+k_2+k_3+k_4)+4)L} \cr
  &\times
{4(k_{4+}-k_{3+})(k_{4-}-k_{3-})(k_{1+}-k_{2+})(k_{1-}-k_{2-}) \over
[i(k_3+k_4-k)+2](k_+^2+1)(k_-^2+1)[i(k_1+k_2+k)+2]} \cr } \eqn\dia $$

\noindent
Here, $k_\pm \equiv {\w \mp k \over 2} $. For $3 \rightarrow 1$ scattering the
on-shell condition is $k_{4-} =0$, and $k_{i+}=0$ for the rest, giving $0$
result
for the above integral. ( One has to make sure that the integral multiplying
this zero factor does not blow up on-shell. It turns out that it has a finite
value.)
Naively saying, this result tells us that the gravitational scattering is $0$!
It is actually not so surprising. For a generic value of momenta, the
gravitational field (including dilaton) has no dynamical degrees of freedom.
Therefore once we
fix the gauge and impose the on-shell conditions, (which is done by looking
for
the residue of the resonance pole in $k_3,k_2,k_1$), its contribution
vanishes!
However, as shall be made especially transparent in Sec. 5, this is not quite
correct. There are specific values of momenta for which the dynamical degree
of
freedom for the
gravity exists. Such values of momenta are called discrete states. In the
conformal gauge these states are killed, meaning that we overfixed
the gauge. An easy way to see this, at the classical level, is to consider the
 equations of motion. Once we fix the conformal gauge, the equations of
motion are:

$$  { \delta S \over \delta G^{+-}}={\delta S \over \delta \Phi }={\delta S
\over \delta T } = 0  \eqn\eom $$

\noindent
For a  generic value of momenta there is no problem since the equation :

$${ \delta S \over \delta G^{\pm\pm}}=0 \eqn\dodo $$ is not independent on
these
\noindent
equations because of the gauge symmetry. Indeed, the gauge symmetry leads to
the
following identities in this gauge:

$$  \eqalign{ (\partial_+ G^{+-}
) {\delta S \over \delta G^{+- }}+ 2 \partial_+ \left( G^{+-} {\delta S \over
\delta G^{+- }} \right)+ 2 \partial_- \left( G^{-+} {\delta S
\over \delta G^{++ }} \right)
 + \left( \partial_+ \Phi \right) {\delta S \over \delta \Phi } +
\left( \partial_+ T \right) {\delta S \over \delta T } &= 0 \cr
(\partial_- G^{-+}
) {\delta S \over \delta G^{-+ }}+ 2 \partial_- \left( G^{-+} {\delta S \over
\delta G^{-+ }} \right)+ 2 \partial_+ \left( G^{+-} {\delta S
\over \delta G^{-- }} \right)
 + \left( \partial_- \Phi \right) {\delta S \over \delta \Phi } +
\left( \partial_- T \right) {\delta S \over \delta T } &= 0 \cr } \eqn\id$$

\noindent
 From \eom~ and \id~ we get

$$ \partial_{\mp} (G^{+-} {\delta S \over \delta G^{\pm\pm} } ) = 0
\eqn\mibun$$

\noindent
Although, for a generic momenta, \mibun~ implies \dodo~, it is obvious there
is
a measure zero degree of freedom in $G_{\pm\pm}$, unjustly killed in the
conformal
gauge, for which \dodo~ cannot be recovered. Indeed, to the leading order, the
equations derived from the gauge-fixed
Lagrangian can be rewritten as:

$$\eqalign{ a_1\varphi &= -{1 \over 8}S^2 \cr
 a_1 e^{-2q} (\partial_+^2-1)(\partial_-^2-1) \zeta  &=  S\partial_+\partial_-
S
-(\partial_+S )(\partial_-S) \cr } \quad {\rm (conformal \ gauge)} \eqn\con
$$

\noindent
On the other-hand, the leading order equation obtained from $(\pm\pm)$
constraint is

$$\eqalign{ a_1\varphi &= -{1 \over 8}S^2 \cr
 a_1e^{-2q} (\partial_+^2-1)(\partial_- -1) \zeta  &=  -S\partial_+ S
-(\partial_+S )^2 \cr
 a_1e^{-2q} (\partial_+ +1)(\partial_- ^2-1) \zeta  &=  -S\partial_- S
+(\partial_-S)^2 \cr}.\quad {\rm (constraint \ equation)} \eqn\pol
$$

\noindent
The equation for $\varphi$ is trivial.
To the leading order, we have that $\partial_+\partial_- S=0$. Also, one can
see
that the equation for $\zeta$ in  \con~ follows from those in \pol~ by taking
one
more derivative, but not the other way round. For $3 \to 1$ scattering, the
incoming wave satisfies $\partial_+ S =0$, and the above equation reduces to

 $$\eqalign{  a_1e^{-2q} (\partial_+^2-1)(\partial_-^2-1) \zeta  &=   0 \cr }
\quad {\rm (conformal \ gauge)} \eqn\cono
$$

\no and

$$\eqalign{  a_1e^{-2q} (\partial_+^2-1)(\partial_- -1) \zeta  &=  0 \cr
 a_1e^{-2q} (\partial_+ +1)(\partial_-^2-1) \zeta  &=  -S\partial_- S
-(\partial_-S)^2 \cr}\quad {\rm (constraint \ equation)} \eqn\polo
$$

\noindent
The equation \polo~ should be used to obtain the correct scattering
contribution. (Feynman diagram result we obtained above corresponds to using
the equation \cono~ instead).

In summary, we expect discrete states to contribute to the gravitational
scattering. In the conformal gauge they can be recovered by relaxing
 the conformal gauge constraints. This resembles the collective field theory
case. There, the naive Feynman diagram vanishes on-shell, but one gets a
nonzero scattering (without poles) after one imposes a Dirichlet boundary
condition [\Deme].
In our case, comparison between \con~ and \pol~ suggests that we should
replace:

$$ { (k_{1+}-k_{2+})(k_{1-} - k_{2-}) \over k_+ + i } \to k_{1-} + k_{2-} + 2i
k_{1-} k_{2-} $$ in \dia~.

\noindent
We then obtain:

$$ \eqalign{ S_\zeta &= {4i \over a_1} \d(k_4-k_3-k_2-k_1) \int dk { k_{4+}
k_{3-}   \over (k+k_1+k_2+2i)(k-k_1-k_2+2i)  } \cr &\times {
(k_{1-}+k_{2-}+2ik_{1-} k_{2-} ) e^{(i(k_1+k_2+k_3+k_4)+4)L}
\over (k-k_1-k_2-2i) (k+k_1+k_2-2i)
(k-k_3-k_4+2i)}} $$

\noindent
where we already put the on-shell condition for the 3 to 1 scattering.
Again, the integral can be evaluated by, say,  making the contour enclose the
upper
half-plane and picking up the poles at $k=k_1+k_2+2i, k=-k_1-k_2+2i$. We have:

$$ \eqalign{ S_\zeta &= -{ \pi \over 4a_1}\d(k_4-k_3-k_2-k_1)  { k_4 k_3
(k_4-k_3+2ik_1 k_2)(k_4+k_3-6i) e^{2(ik_4+2)L} \over
(k_4-k_3+2i)(k_4-k_3-2i)(k_4-2i)(k_3-2i)}
\cr
&\sim {\pi \over a_1} \delta(k_4-k_3-k_2-k_1) { -k_2^2   \over
(k_4-2i)(k_3-2i)}\cr } $$

\noindent
In the end, of course, we have to sum over the inequivalent permutations of
external momenta.
One can easily check that this result agrees with the pole contribution
extracted from the string amplitude \strS~, if one fixes $a_1 = 1/2$.
Let us note in passing that the values of $a_1,a_2,b_1,b_2$ obtained here are
in
accordance with [\Natsu].
 Although
the detailed calculations will be not presented here, we have also
 evaluated the
two-point function to the next order in $\mu$ and found that the quartic
tachyon interaction vertex of the form $e^{4q} T^4$ should vanish in
 order to reproduce
the correct pole structure.

If we fix some other gauges, such as Lorentz or Siegel's gauge, the
contributions
from the discrete states are not killed and we expect to get the correct $S$
matrix without imposing additional constraints. It can be easily done using
the
Feynman diagram calculations, but instead of repeating the same kind of
calculations, we will turn to these other gauges in the context of the string
field theory in later sections. The nature of discrete states will be
described
in more detail in section 5 when
we consider Lorentz gauge for the string field theory.

\chapter{\bf Gauge Invariant Action and $2d$ Closed String Field Theory}
In the previous section we presented a detailed account on how
 to
 perform perturbative calculations in the effective Lagrangian approach.
 An important property of the formalism is that it has a
 non-trivial cosmological constant naturally built-in (although
 performing the actual calculations for the higher powers in $\mu$ may
 be very cumbersome). On the other hand, BRST formulation of the CSFT
corresponds to $\mu = 0$. This apparent drawback is compensated,
 however, by the fact that the contributions from discrete states (gauge
states
in $2d$) are captured much more naturally within the CSFT framework.
 It is our task in the remainder of the paper to discuss the CSFT in relation
with the effective Lagrangian results of Sec. 2.

In this section we consider the gauge invariant formulation of the
 theory. Before we start, let us make some comments on our approach and
 the formalism we use. Throughout, we favor simplicity. As a rule, we
 observe certain properties on the first couple of levels and then
 generalize the results to an arbitrary high level. In this way we hope
 to build a solid intuitive picture for what is going on. To
 consistently describe BRST closed string field theory in $2d$ one
 needs, in principle, to include the kinetic term (which contains
 the BRST charges $Q$, $\bar{Q}$), cubic, quartic, and all the
 higher order interactions [\Kaku].
Fortunately, for our purposes, namely to solve the free EOM, and
 to calculate the $4$-point tachyon--tachyon scattering amplitudes,
 it is enough to know only $\bra{V_2}$ and $\bra{V_3}$ vertices. In
 that case, the gauge invariant action reads, approximately:

$$
S_{g. \, i.} = \, {1 \over 2} \bra{\Phi} \, c_0^{(-)} \,
(Q + \bar{Q}) \, \ket{\Phi} \, + \, {1 \over 3}
\bra{V_3} \ket{\Phi}_1  \ket{\Phi}_2 \ket{\Phi}_3 \eqno\eq
$$

\noindent
Note that from this section on we use slightly different notation than in Sec.
2
(see [\Uro]). As mentioned above, the two are related by a Wick rotation of
the
momenta, and a rescaling by the factor of $\sqrt 2$. Vertices in $(3.1)$ are
direct products of the corresponding open string vertices.  A string field
$\ket{\Phi}$ must satisfy the conditions
 $(L_{0} - \bar{L}_0) \ket{\Phi} = 0$, and $b_0^{-} \ket{\Phi} = 0$.
 Our conventions for the ghost number counting are as follows. Ghost
 zero modes $c_0^{\pm} \e {c_0^{+} \pm c_0^{-} \over 2}$ and $b_0^{\pm}
 \e b_0^{+} \pm b_0^{-}$ have ghost numbers $\pm 1$, respectively,
 while string field $g(\ket{\Phi}) = 2$. Since the scalar product
 needs $6$ ghost zero modes to saturate, $g(Q) = g(\bar{Q}) = 1$.

A generic string field, subject to the restrictions described above,
 can be expanded in terms of the coefficient functions in the
 following way:

$$  \eqalign
{
\ket{\Phi} =& \, \ket{p} \phi(p) \, + \, \, \am_{-1}
 \bar{\alpha}^{\nu}_{-1} \ket{p}  h_{\mu \nu} (p) \, + \,
 b_{-1} \bar{c}_{-1} \ket{p} \psi(p) \, - \, c_{-1} \bar{b}_{-1} \,
 \ket{p} \, \eta(p) +  \cr
&+ \,  c_0^{+}(\, \am_{-1} \bar{b}_{-1} \ket{p} \, i A_{\mu}(p)
 \, + \, b_{-1} \bar{\alpha}^{\mu}_{-1} \ket{p} \,
 i B_{\mu}(p) \, ) \, , \cr
}
\eqno\eq
$$

\noindent
In this section we keep only up to the first level excitations.
 The quadratic part of the action $(3.1)$ then reads:

$$   \eqalign
{
S^{(2)} =& \int d^2 x \, e^{i Q \cdot x} (\,  {1 \over 2}
 \partial \phi \cdot \partial \phi \, - \phi^2 \, +
 {1 \over 2} \partial h_{\mu \nu} \partial h_{\mu \nu}
 \, - \partial \psi \partial \eta + \, \cr
&+ \, 2 A_{\mu}^2 + \, 2 A^{\mu} (D_{\nu} h_{\mu \nu}
 + \, \partial_{\mu}\psi) \, + \, 2 B_{\mu}^2 + \, 2 B^{\mu}
 (D_{\nu} h_{\nu \mu} + \, \partial_{\mu} \eta) \, ) \, \cr
}
\eqno\eq
$$

\noindent
Here and below we use the shorthand $D_{\mu}$ to denote the covariant
 derivative $D_{\mu} = \partial_{\mu} + \, i Q_{\mu}$.

Let us now discuss the gauge transformations. At the linearized level
 they are given by:

$$   \eqalign
{
&\delta \ket{\Phi} = \, (Q + \bar{Q}) \, \ket{\Lambda}
  \, , \cr
&b_0^{+} \,  \ket{\Lambda} = \, L_0^{+} \,
  \ket{\Lambda} = \, 0
}
\eqno\eq
$$

\noindent
Here, gauge parameter field $\ket{\Lambda}$ has a ghost number $g=1$
 (remember that the matter field has $g=2$). The component expansion
 for a gauge field, up to the first level, is:

$$
 \ket{\Lambda} = \, \am_{-1} \bar{b}_{-1} \, \ket{0} \, i a_{\mu}(x)
 \, + \, \bar{\am}_{-1} b_{-1} \, \ket{0} \, i b_{\mu}(x) \, + \,
 c^{+}_0 \, b_{-1} \bar{b}_{-1} \ket{0} \sigma(x)
\eqno\eq
$$

\noindent
 From $(3.5)$, using $(3.4)$, one can easily read--off the gauge
 transformation for the component fields:

$$  \eqalign
{
&\delta \phi(x) = \, 0 \, , \cr
&\delta h_{\mu \nu}(x) = \, \partial_{\nu} a_{\mu} + \,
 \partial_{\mu} b_{\nu} \, ,\cr
&\delta \psi(x) = \, - \, D_{\mu} b_{\mu} \, - 2 \sigma, \cr
&\delta \eta(x) = \, - \, D_{\mu} a_{\mu} \,
  + 2 \sigma  \, ,\cr
&A_{\mu} = \, - {1 \over 2} \partial \cdot D \, a_{\mu}
 \, + \, \partial_{\mu} \sigma \, , \cr
&B_{\mu} = \, - {1 \over 2} \partial \cdot D\, b_{\mu}
 - \, \partial_{\mu} \sigma \, , \cr
}
\eqno\eq
$$

\noindent
It is easy to check that the transformations $(3.6)$ indeed leave the action
$(3.3)$ invariant.

To be able to compare the action $(3.3)$ with the effective Lagrangian $(2.1)$,
we need to
obtain an alternative
 form of $(3.3)$.
To this end, note that in general $h_{\mu \nu}$ is not a symmetric tensor.
This
 is reflected in the fact that the left and the right indices
 transform differently under $(2.6)$ (one through $a_{\mu}$ and
 the other one through $b_{\nu}$). Notice, also,  that  $A_{\mu}$ and
$B_{\mu}$,
the coefficients functions
 in front of $c_0^{+}$, are the Lagrange multipliers. As such,
 they are not dynamical and can be integrated out using the corresponding EOM:

$$  \eqalign
{
A_{\mu} =& \, - {1 \over 2} (\, D_{\nu} h_{\mu \nu} + \,
 \partial_{\mu} \psi)  \, \cr
B_{\mu} =& \, - {1 \over 2} (\, D_{\nu} h_{\nu \mu} + \,
 \partial_{\mu} \eta)  \, \cr
}
\eqno\eq
$$

\noindent
The action $(3.3)$ then becomes:

$$   \eqalign
{
S^{(2)} =& \int d^2 x \, e^{i Q \cdot x} (\,  {1 \over 2}
 \partial \phi \cdot \partial \phi \, - \phi^2 \, + {1 \over 2}
 \partial h_{\mu \nu} \partial h_{\mu \nu} \, - \partial \psi
 \partial \eta + \, \cr
& -{1 \over 2} \, (D_{\nu} h_{\mu \nu} + \, \partial_{\mu}\psi)^2
  \, - {1 \over 2}(D_{\nu} h_{\nu \mu} + \,
 \partial_{\mu} \eta)^2 \, \cr
}
\eqno\eq
$$

\noindent
One can simplify the action even further by identifying $\psi = \eta$.
 In that case, the field $\psi$ is called dilaton. The action
 $(3.8)$ becomes:

$$   \eqalign
{
S^{(2)} =& \int d^2 x \, e^{i Q \cdot x} (\,  {1 \over 2} \partial
 \phi \cdot \partial \phi \, - \phi^2 \, + {1 \over 2} \partial
 h_{\mu \nu} \, \cdot \partial h_{\mu \nu} \, - 2 \, \partial
 \psi \, \cdot \partial \psi  \, \cr
& - {1 \over 2} \partial_{\chi} h_{\nu \mu} \,
 \partial_{\nu} h_{\chi \mu} - \, {1 \over 2}
 \partial_{\chi} h_{\nu \mu} \,  \partial_{\nu} h_{\mu \chi}
 \, +(h_{\nu \mu} + h_{\mu \nu}) \, \partial_{\mu}
 \partial_{\nu} \, \psi ) \cr
}
\eqno\eq
$$

\noindent
Note that the tensor $h_{\mu \nu}$ can be decomposed into a symmetric
 and antisymmetric parts $h_{\mu \nu}(p) = \, g_{\mu \nu}(p) +
\epsilon_{\mu \nu} j(p)$, and that dilaton couples only to the
 symmetric part $g_{\mu \nu}$.
The action $(3.9)$, in fact, does not depend on $j$:

 $$   \eqalign
{
S^{(2)} =& \int d^2 x \, e^{i Q \cdot x} (\,  {1 \over 2}
 \partial \phi \cdot \partial \phi \, - \phi^2 \, + {1 \over 2}
 \partial g_{\mu \nu} \, \cdot \partial g_{\mu \nu} \, - 2 \,
 \partial \psi \, \cdot \partial \psi  \, \cr
& - \partial_{\chi} g_{\nu \mu} \, \partial_{\nu} g_{\chi \mu}
  \, + \, 2 \, g_{\mu \nu} \, \partial_{\mu}
 \partial_{\nu} \, \psi ) \cr
}
\eqno\eq
$$

\noindent
Since in a symmetric case we can identify $a_{\mu} = b_{\mu}$,
the gauge transformations $(3.6)$ simplify as well:

$$   \eqalign
{
&\delta \phi(x) = \, 0 \, , \cr
&\delta h_{\mu \nu}(x) = \, \partial_{\nu} a_{\mu} + \,
 \partial_{\mu} a_{\nu} \, ,\cr
&\delta \psi(x) = \, - \, D_{\mu} a_{\mu} \, , \cr
}
\eqno\eq
$$

\noindent
Careful comparison with the effective Lagrangian of the last
section
shows that the latter can
be put into the form $(3.10-3.11)$ by keeping just the quadratic terms.
To that end, if one makes a transformation $ \psi =  f - {1 \over 2}  g_{\m\m}
$
and plugs it into $(3.10)$, one obtains:

$$\eqalign{ S^{(2)} &= \int d^2x e^{iQ \cdot x} \Big( {1 \over 2} \partial
g_{\mu
\nu} \cdot  \partial  g_{\mu \nu} -  \partial_\chi  g_{\nu
\mu} \partial_\nu  g_{\chi \mu} \cr
&- g_{\m\n} \partial_\m \partial_\n  g_{\a\a}
-{1 \over 2} \partial  g_{\a\a} \cdot \partial  g_{\b\b} - 2\partial
 f \cdot \partial f \cr
&+ 2  g_{\m\n} \partial_\m\partial_\n  f + 2 \partial  g_{\a\a}
\cdot \partial  f \cr
&+{1 \over 2} \partial \phi \cdot
\partial \phi -  \phi^2 \Big) \cr } \eqn\fie $$

\noindent
On the other hand, expanding in \dil~ around the linear dilaton
vacuum, $ G_{\mu\nu} = \eta_{\mu\nu} + g_{\mu\nu}, \Phi = 2q + f , $ we
see that the quadratic part of the Lagrangian $(2.1)$ can be brought to the
form:

$$\eqalign{ S^{(2)} &={a_1 \over 2} \int d^2x e^{-4q} \Big[ -{1
\over 4} \partial g_{\mu
\nu} \cdot  \partial g_{\mu \nu} +{1 \over 2}  \partial_\chi g_{\nu
\mu} \partial_\nu g_{\chi \mu} \cr
&+{1 \over 2} g_{\m\n} \partial_\m \partial_\n g_{\a\a}
+{1 \over 4} \partial g_{\a\a} \cdot \partial g_{\b\b} + 4\partial f \cdot
\partial f \cr
&- 2 g_{\m\n} \partial_\m\partial_\n f - 2 \partial g_{\a\a} \cdot \partial f
\Big] \cr
&+{1 \over 2} \int d^2x e^{-4q} \left[ -(\partial T)^2 + 4 T^2
\right]
\cr } \eqn\simp $$

\noindent
after appropriate partial integrations. Here and below, the repeated indices
are
summed over. We immediately see that \fie~ and \simp~ identical up to renaming
of some fields and coordinates.

In principle, one
 can perform similar analysis for the higher level states as well.
 However, the gauge invariant action is, in that case, too
 cumbersome to be illuminating. Still, it might be of interest
 to try to simplify it as much as possible using the symmetry arguments
 similar to the above, since that would provide us, possibly, with
 a hint on how to generalize the effective Lagrangian to the higher level
excitations.

\chapter{{\bf Gauge Fixing -- Siegel's Gauge}}

In this section we consider the gauge fixing in Siegel's gauge. It is
 the most commonly used covariant gauge in string field theory. It
 is especially convenient, as we shall observe later, for the
 calculations of scattering amplitudes in the field--theoretic
 framework. To show in more detail the formalism at work, we
 consider the excited states on both the first and the second levels.

To impose Siegel's gauge, the following condition on a string field
 must be satisfied:

$$
b_{0}^{+} \ket{\Phi} = 0
\eqno\eq
$$

\noindent
It is well-known that in order to impose the condition $(4.1)$,
 one has to introduce ghosts--of ghosts -- ... of ghosts, using,
 say, the Batalin--Vilkovisky formalism (this is because the theory
 has an infinite degree of reducibility). In this paper we are
 interested only in $g = 2$ physical states, and are calculating
 tree level amplitudes exclusively, so we do not have to worry
 about the ghosts of ghosts (see [\Uros] for more details)

Let us illustrate the effect of the condition $(4.1)$ on the
 level-one free action. Since $\{ b_0^{\pm} , c_0^{\pm} \} = \, 1$,
 and $b_0^{\pm} \ket{p} = 0$, the net effect of imposing $(4.1)$
 is equivalent to requiring that $A_{\mu} = B_{\mu} = 0$. In
 Siegel's gauge, all Lagrange multipliers are, therefore, set to zero. The
gauge-fixed version of the action $(3.3)$ is, thus:

$$
S^{(2)}_{g.f.}  = \, \int d^2 x \, e^{i Q \cdot x} (\,  {1 \over 2}
 \partial \phi \cdot \partial \phi \, - \phi^2 \, + {1 \over 2}
 \partial h_{\mu \nu} \partial h_{\mu \nu} \,  - \partial \psi \partial \eta
\,
)
\eqno\eq
$$

\noindent
and the gauge conditions are:
$$  \eqalign
{
D_{\nu} h_{\mu \nu} + \, \partial_{\mu} \psi = \, 0 \, , \, \, \,
D_{\nu} h_{\nu \mu} + \, \partial_{\mu} \eta = \, 0 \, .
}
\eqno\eq
$$

\noindent
Gauge conditions $(4.3)$ can be used to simplify the action $(4.2)$
 by eliminating $\psi$ and $\eta$ fields:

$$
S^{(2)}_{g.f.}  = \, \int d^2 x \, e^{i Q \cdot x} (\,  {1 \over 2}
 \partial \phi \cdot \partial \phi \, - \phi^2 \, + {1 \over 2}
 \partial h_{\mu \nu} \partial h_{\mu \nu} \,  - \partial_{\nu}
 h_{\mu \nu} \partial_{\chi} h_{\chi \mu} \,)
\eqno\eq
$$

\noindent
 From $(4.4)$ it is clear that in Siegel's gauge only tachyon and graviton
 can be expected to `propagate', the fact which is confirmed in Sec. 6,
 where we show that the potential residue from the dilaton intermediate
 state in the scattering amplitude vanishes.

Let us now consider second level contributions. A field satisfying
 the conditions $L_{0}^{-} \ket{\Phi} = \, b_{0}^{-} \ket{\Phi} = \, 0$
 has the following generic form:

$$
\ket{\Phi} = \ket{\Phi}_{b_0^{+} = 0} + \, \ket{\Phi}_{b_0^{+}
 \ne 0} \, .
$$

\noindent
On the second level, there is $17$ fields satisfying Siegel's gauge condition:

$$  \eqalign
{
\ket{\Phi}_{b_0^{+} = 0}=& \, \am_{-1} \, \an_{-1} \,
\bar{\alpha}_{-1}^{\chi} \, \bar{\alpha}_{-1}^{\la} \, \ket{p}
 A_{\mu \nu \chi \la} \, + \, \am_{-1} \, \an_{-1} \,
 \bar{\alpha}_{-2}^{\chi} \, \ket{p} B_{\mu \nu \chi}  \, \cr
&+ \, \bar{\am}_{-1} \, \bar{\an}_{-1} \,
 \alpha_{-2}^{\chi} \, \ket{p} C_{\mu \nu \chi} \, + \,
\alpha^{\mu}_{-2} \, \bar{\alpha}^{\nu}_{-2} \ket{p}
 \, D_{\mu \nu} \, + \, b_{-1} \, c_{-1} \bar{\alpha}_{-1}^{\mu} \,
\bar{\alpha}_{-1}^{\nu} \, \ket{p} E_{\mu \nu} \, +  \, \cr
&+ \, \, \bar{b}_{-1} \, \bar{c}_{-1} \alpha_{-1}^{\mu} \,
 \alpha_{-1}^{\nu} \, \ket{p} F_{\mu \nu} \, + \,
  \bar{b}_{-1} \, c_{-1} \bar{\alpha}_{-1}^{\mu} \,
 \alpha_{-1}^{\nu} \, \ket{p} G_{\mu \nu} \, +  \cr
&+ \, b_{-1} \, \bar{c}_{-1} \alpha_{-1}^{\mu} \,
 \bar{\alpha}_{-1}^{\nu} \, \ket{p} H_{\mu \nu} \, + \,
 i b_{-1} \, c_{-1} \bar{\alpha}_{-2}^{\mu} \ket{p} J_{\mu} \, +  \, \cr
&+ \, i \bar{b}_{-1} \, \bar{c}_{-1} \alpha_{-2}^{\mu}
 \ket{p} K_{\mu} \, + \, b_{-1} \, c_{-1} \, \bar{b}_{-1} \,
\bar{c}_{-1} \ket{p} \, L \, + b_{-2} \bar{c}_{-2} \, \ket{p}
 \Upsilon \, + \cr
&+ \, \bar{b}_{-2} \, c_{-2} \, \ket{p} \Psi \, + \, i
 \bar{b}_{-1} \, c_{-2} \, \bar{\alpha}^{\mu}_{-1} \,
\ket{p} \Theta_{\mu} \, + \, i b_{-1} \, \bar{c}_{-2} \, \alpha^{\mu}_{-1} \,
\ket{p} \Gamma_{\mu} \cr
&+ \, i c_{-1} \, \bar{b}_{-2} \, \alpha_{-1}^{\mu}
 \ket{p} X_{\mu} + \,  i \bar{c}_{-1} \, b_{-2} \,
 \bar{\alpha}_{-1}^{\mu} \ket{p} Y_{\mu} + \, \cr
}
\eqno\eq
$$

\noindent
and $12$ fields which violate it:

$$  \eqalign
{
\ket{\Phi}_{b_0^{+} \ne 0}=& \, c_0^{+}(\, i \, \bar{b}_{-1} \,
 \am_{-1} \, \an_{-1} \, \bar{\alpha}^{\chi}_{-1} \, \ket{p}
 m_{\mu \nu \chi} \, + \, i \, b_{-1} \, \bar{\am}_{-1} \,
 \bar{\an}_{-1} \, \alpha^{\chi}_{-1} \, \ket{p} n_{\mu
 \nu \chi} \, + \, \cr
&+ \, \bar{b}_{-1} \, \bar{\alpha}^{\mu}_{-1} \, \alpha^{\nu}_{-2}
 \, \ket{p} o_{\mu \nu} \, + \, b_{-1} \, \alpha^{\mu}_{-1} \,
\bar{\alpha}^{\nu}_{-2} \, \ket{p} s_{\mu \nu} \, + \, \cr
&+ \, i \, b_{-1} \, \bar{b}_{-1} \, \bar{c}_{-1} \,
 \alpha^{\mu}_{-1} \, \ket{p} \, r_{\mu} \, + \, i \, \bar{b}_{-1}
 \, b_{-1} \, c_{-1} \, \bar{\alpha}^{\mu}_{-1} \, \ket{p}
 \, t_{\mu} \, \cr
&+ \, b_{-2} \, \bar{\alpha}^{\mu}_{-1} \, \bar{\alpha}^{\nu}_{-1}
 \, \ket{p} \, q_{\mu \nu} \, + \, \bar{b}_{-2} \, \alpha^{\mu}_{-1} \,
\alpha^{\nu}_{-1} \, \ket{p} \, p_{\mu \nu} \, \cr
&+ \, i \, b_{-2} \, \bar{\alpha}_{-2}^{\mu} \, \ket{p} \,
\beta_{\mu} + \, i \, \bar{b}_{-2} \, \alpha^{\mu}_{-2} \,
 \ket{p} \gamma_{\mu} \, + \, \cr
&+ \, b_{-2} \, \bar{b}_{-1} \bar{c}_{-1} \, \ket{p} \,
 \delta \, + \, \bar{b}_{-2} \, b_{-1} \, c_{-1} \, \ket{p}
 \omega  \,) \, \,  ,\cr
}
\eqno\eq
$$

\noindent
It is quite clear that a gauge invariant action on that level (and,
 to that matter, on all the levels above it) is a very large expression.
 In fact, it consists of $132$ terms. As we have mentioned above,
 one could, in principle, try to simplify the action by symmetry
 arguments, and by exploiting the fact that fields $(4.6)$ are all
 Lagrange multipliers. We shall not pursue that here, however,
 turning, instead, directly to the gauge fixing. It turns out that it
simplifies
 matters a lot. In fact, in Siegel's gauge all Lagrange multipliers are
 taken to vanish, which gives $12$ tensorial gauge conditions on
 the remaining fields. The gauge fixed action reads:

$$  \eqalign
{
S_{g.f} =& \, {1 \over 2} \, \bra{\Phi} \, c^{-}_0 \, c^{+}_0 \,
 L^{+}_0 \, \ket{\Phi} = \, \int \, e^{i Q \cdot x} \, (\, 2
 \partial A_{\mu \nu \chi \la} \, \partial A_{\mu \nu \chi \la}
 + \, 4 A_{\mu \nu \chi \la} \, A_{\mu \nu \chi \la} \, + \, \cr
&+ \, 2 \partial B_{\mu \nu \chi} \, \partial B_{\mu \nu \chi}
 \, + \, 4 \, B_{\mu \nu \chi} B_{\mu \nu \chi} + \, 2 \partial
 C_{\mu \nu \chi} \, \partial C_{\mu \nu \chi} \, + \, 4
 \, C_{\mu \nu \chi} C_{\mu \nu \chi} \, + \cr
&+ \, 2 \partial \, D_{\mu \nu} \, \partial \, D_{\mu \nu} \, + \,
4 D_{\mu \nu} \, D_{\mu \nu} - \, \partial \, E_{\mu \nu} \,
 \partial \, E_{\mu \nu} \, - \, 2 \, E_{\mu \nu} E_{\mu \nu} \,
 - \,  \cr
&- \, \partial \, F_{\mu \nu} \, \partial \, F_{\mu \nu} \, - \, 2
\, F_{\mu \nu} F_{\mu \nu} \, - \, \partial \, G_{\mu \nu} \, \partial
 \, H_{\mu \nu} \, - \, 2 \, G_{\mu \nu} H_{\mu \nu} \, - \, \cr
&- \, \partial \, J_{\mu} \, \partial \, J_{\mu} \, - \, 2 \,
J_{\mu} J_{\mu} \, - \, \partial \, K_{\mu} \, \partial \,
 K_{\mu} \, - \, 2 \, K_{\mu} \,  K_{\mu}\, + \, \cr
&+ \, {1 \over 2} \, \partial \, L \, \partial \, L \, + \,
 L^2 \,  - \, \partial \Psi \, \partial \Upsilon \, - 2 \Psi
 \Upsilon \, \cr
&+ \, \partial \Theta_{\mu}  \partial Y_{\mu} \, + \, 2
 \Theta_{\mu} Y_{\mu} \, + \, \partial \Gamma_{\mu}  \partial
 X_{\mu}  \, + \, 2 \Gamma_{\mu} \, Y_{\mu} \cr
}
\eqno\eq
$$

\noindent
This is the expression which we shall use in Sec. 6 to determine the
 propagators
of the fields. Just like on the level-one case, we shall observe that
 this is not the minimal expression for the gauge-fixed action, and
 that, in fact, only some of fields present in the action $(4.7)$
  contribute to scattering.

\chapter{\bf Lorentz Gauge and Physical States}

This section is devoted to the discussion of the physical states in $2d$
closed
string field theory(we closely follow [\Poly]). Here, a remark is in order. On
general grounds one expects
that in $2d$ gauge theory of arbitrary spin there should be no physical
degrees
of freedom. On the other hand, above we have witnessed an abundance of fields
entering the $2d$ string field theory. The natural question arises: how to
reconcile these two facts? In attempt to answer this question, we take a
straightforward route. Namely, we try to ``gauge-out''
as much fields as possible. In an ideal world, in the end of the day we would
be
left with a tachyon (the ground state $\phi$), whereas all of the gauge fields
would have to be gauged out. In reality, this is almost the case for the
closed
$2d$ string theory. The only remnants of the gauge fields are the so--called
discrete states.
In this section our primary purpose is to discuss the origin of the discrete
states from the field--theoretic point of view. To locate the physical states
it
proves advantageous to use the Lorentz-like gauge. We shall discuss in detail
the
physical states on the first two levels, from which it should be clear how the
procedure generalizes to an arbitrary high level.

On the first level the action, in its most general gauge invariant form, is
given by the Eq. $(3.3)$, from which one easily derives the following
equations
of motion for the component fields (they correspond to the Eq.
 $(Q + \bar{Q}) \ket{\Phi}$):

$$   \eqalign
{
&\partial \cdot D h_{\mu \nu} + \, 2 \partial_{\nu} A_{\mu} + 2 \partial_{\mu}
B_{\nu} = \, 0  \, \cr
&\partial \cdot D \eta \, - \, 2 D_{\mu} A_{\mu} =  \, 0 \, , \cr
&\partial \cdot D \psi \, - 2 D_{\mu} B_{\mu} = \, 0 \, , \cr
&2 A_{\mu} + \, D_{\nu} \, h_{\mu \nu} + \, \partial_{\mu} \psi \, = 0 \, ,
\cr
&2 B_{\mu} + \, D_{\nu} \, h_{\nu \mu} + \, \partial_{\mu} \eta \, = 0 \, .
\cr
}
\eqno\eq
$$

\noindent
We can now use the gauge transformations $(3.6)$ to fix the gauge in which
$D_{\nu} h_{\mu \nu} = \psi =  \eta = 0$. This gauge, which we refer to as
Lorentz gauge, has an advantage that in it, EOM take a particularly
simple form. Namely, we see that the only potentially nontrivial field is
field
$h_{\mu \nu}$ (as expected). It satisfies the following on--shell and
polarization conditions (in the momentum space):

$$  \eqalign
{
&p \cdot (p+Q) h_{\mu \nu} = \, 0  \, ,  \cr
&(p+Q)_{\nu} h_{\mu \nu} = 0  \, .   \cr
}
\eqno\eq
$$

\noindent
We are looking for the non-trivial solutions of the Eq. $(5.2)$ using the
ansatz
$h_{\mu \nu} = \epsilon_{\mu} \bar{\epsilon}_{\nu}$. In other words, we treat
left and the right moving sectors independently. One obvious solution to
$(5.2)$
is $\epsilon_{\mu} = p_{\mu} \la$, for an arbitrary $\la$ (and a $p_{\mu}$
satisfying the on--shell condition $(p \cdot (p+ Q) = \, 0$). But this is
nothing but a gauge transformation. Such a solution is, indeed, trivial.
However, there are special values of momenta for which either gauge
transformations do not exist ($p_{\mu} = 0$), or the gauge conditions relax,
since they are satisfied for any coefficient function ($p_{\mu} = - Q_{\mu}$).
In either case, there is a non-trivial solution which reads:

$$
h^{(0)}_{\mu \nu} = \, \epsilon_{\mu \chi} Q_{\chi} \epsilon_{\nu \la} Q_{\la}
\, .
\eqno\eq
$$

\noindent
One can immediately see that this solution is incompatible with the conformal
gauge
condition $G_{\pm \pm} = 0$. It is not surprising, therefore, that the
contribution from this state to
the scattering amplitude is killed in the conformal gauge (see Sec. 2.2).
Suitably normalized Fock space states corresponding to the
solution $(5.3)$ are:

 $$
\alpha^m_{-1} \bar{\alpha}^m_{-1} \, \ket{0, 0} \, , \, \,  \alpha^m_{-1}
\bar{\alpha}^m_{-1} \, \ket{0, - 2 {\sqrt 2}}   \, ,
\eqno\eq
$$

\noindent
where by superscript $m$ we have denoted matter oscillators. It is important
to
note that in this gauge only matter field contributes to excitations. In the
literature states $(5.4)$ are denoted as $W^{(\pm)}_{1 ,0}$ [\Witt]. These
states are all what is left from the graviton after the gauge fixing!

One can follow the same procedure on the next level. Of course, the
 fact that one starts with $29$ fields could make calculations quite
 involved. One should, therefore, try to simplify things as
much as one can from the outset, using the already accumulated information.
First of all, $12$ of these fields are Lagrange multipliers (see Eq.
$(4.10)$).
They, just like fields $A_{\mu}$ and $B_{\mu}$ on the firsts level, enter the
equations of motion $(Q + \bar{Q}) \ket{\Phi}$ without derivatives. At this
stage, we can employ
gauge transformations $(3.6)$ to require that:

$$  \eqalign
{
&E_{\mu \nu} = \, F_{\mu \nu} = \, G_{\mu \nu} = \, H_{\mu \nu}  = \,   \, \cr
&J_{\mu} = \, K_{\mu} = \, \Theta_{\mu} = \, \Gamma_{\mu} = \, X_{\mu} = \,
Y_{\mu} = \,     \cr
&L = \,  \Upsilon = \, \Psi = 0  \, . \cr
}
\eqno\eq
$$

\noindent
In addition, we impose the polarization conditions on the remaining
fields $A_{\mu \nu \chi \la}$, $B_{\mu \nu \chi}$,  $C_{\mu \nu \chi}$, and
$D_{\mu \nu}$ (it is from them that physical states remain, so we refer to
them
as `physical fields'). These conditions are:

$$   \eqalign
{
&(p + Q)_{\la} A_{\mu \nu \chi \la} + \, i \, B_{\mu \nu \chi} =  \, (p +
Q)_{\la} A_{\la \mu \chi \nu} + \, i \, C_{\nu \chi \mu} = \, 0 \, ,\cr
&i \, (p + Q)_{\la} C_{\la \mu \nu} + \,  D_{\nu \mu} =  \, i \, (p + Q)_{\la}
B_{\la \mu \nu} + \,  D_{\mu \nu} = \, 0 \, ,\cr
& A^{\la}_{\la \mu \nu} \, + \,  i (2 p + 3 Q)_{\la} C_{\mu \nu \la} = \,
A^{\la}_{\mu \nu \la} \, + \, i (2 p + 3 Q)_{\la} B_{\mu \nu \la} = 0 \, , \cr
&- i \, B^{\la}_{\la \mu} \, + \,  (2 p + 3 Q)_{\la} D_{\la \mu} = \, - i \,
C^{\la}_{\la \mu} \, + \,  (2 p + 3 Q)_{\la} D_{\mu \la} = 0 \, . \cr
}
\eqno\eq
$$

\noindent
Gauge conditions $(5.6)$ ensure that all Lagrange multipier fields $(4.10)$
vanish. Finally, fields must satisfy the EOM which, in this gauge, are simply
the mass shell conditions:

$$  \eqalign
{
&(\, (p + {Q \over 2})^2 + 4 \, ) A_{\mu \nu \chi \la} = \, 0 \, , \cr
&(\, (p + {Q \over 2})^2 + 4 \, ) B_{\mu \nu \chi \la} =  \, (\, (p + {Q \over
2})^2 + 4 \, ) \, C_{\mu \nu \chi} = \, 0 \, , \cr
&(\, (p + {Q \over 2})^2 + 4 \, ) D_{\mu \nu} = \, 0 \, .
}
\eqno\eq
$$

\noindent
All this is very similar to the level-one case. Namely, we have first found
the
EOM for the component fields and identified the Lagrange multipliers. Then, we
have  used the gauge
freedom to fix a particularly simple gauge (Lorentz gauge) in which the
analogues of the fields $\eta$ and $\psi$ are taken to vanish. These are the
coefficient functions corresponding to the states with ghost excitations.
Finally, we have imposed the polarization conditions on the remaining physical
fields (analogous to the graviton field $h_{\mu \nu}$), which made Lagrange
multiplier fields vanish as well. In this way, we got rid of all the fields
but
the set of physical fields.
What remains are discrete states corresponding
to the values of momenta for which either the gauge transformations or the
polarization conditions degenerate.

 Just as before, in order to identify the physical states,
one has to find non-trivial solutions to the EOM $(5.7)$ subject to the
constraints $(5.6)$. To find the solutions, let us, again, use the
decomposition
into
the right and left movers:

$$  \eqalign
{
&A_{\mu \nu \chi \la} = \, R_{\mu \nu} \bar{R}_{\chi \la}  \, \cr
&B_{\mu \nu \chi} = \, R_{\mu \nu} \bar{S}_{\chi}  \, \cr
&C_{\mu \nu \chi} = \, \bar{R}_{\mu \nu} S_{\chi}  \, \cr
&D_{\mu \nu} = \, S_{\mu} \bar{S}_{\nu}  \, \cr
}
\eqno\eq
$$

\noindent
Plugging this ansatz back to the original set of equations $(5.6)$, one
obtains:

$$   \eqalign
{
&(p + Q)_{\mu} \, R_{\mu \nu} + \, i \, S_{\nu} = \, 0 \, , \cr
&- i  \,  \, R^{\mu}_{\mu}  + \, (2 p + 3 Q)_{\mu} S_{\mu} = \, 0  \, ,\cr
}
\eqno\eq
$$

\noindent
with an identical set of equations for the barred variables (and, of course,
the
set of mass--shell conditions which are satisfied if and only if $p \cdot (p +
Q) = - 2$). It is worth noting,
at this point, that we have achieved a substantial simplification. In essence,
instead of trying to solve a system $29$ linear equations, we have to solve
just
two. There are two independent solutions for the constraint equations. One is:

$$  \eqalign
{
&R_{\mu \nu} = {i \over 2} (p_{\mu} \eta_{\nu} + p_{\nu} \eta_{\mu}), \, \cr
&S_{\mu} = \, \eta_{\mu}   \, , \cr
}
\eqno\eq
$$

\noindent
where $\eta_{\mu}$ satisfies $(p+ Q) \cdot \eta = 0$, and the other is:

$$  \eqalign
{
&R_{\mu \nu} =  \, i (\delta_{\mu \nu} + 3 p_{\mu} p_{\nu})  \, \cr
&S_{\mu} = \, (5 p - Q)_{\mu}   \, .
}
\eqno\eq
$$

\noindent
Not surprisingly, both of these solutions correspond to a gauge state: first
to
$L_{-1} (\eta \cdot \alpha_{-1} \ket{p}$, and the second one to $(L_{-2} + \,
{3
\over 2} {(L_{-1})}^2 ) \, \ket{p}$. The corresponding physical states are
$W^{(\pm)}_{{3 \over 2},{\pm {1 \over 2}}}$, in the standard
notation. To complete the story, one needs to take into account also the
antiholomorphic sector (states at $g=2$ which we discuss here are just a
simple
tensor product of the two sectors). We have shown, therefore, that
out of $29$ fields which naively exist in the theory on the second level,
only $4 \times 4 = 16$ discrete states are physical.

It is now rather clear that similar is true for an arbitrary higher level as
well. Out of a
potential abundance of fields on the higher levels, only discrete states are
physical. Two types of discrete states correspond to the `positively dressed
states', for which gauge transformations degenerate, and `negatively dressed'
ones, for which gauge conditions relax. One can always choose a gauge in which
only matter oscillators contribute to the excitations. Physical states are
product of states in the holomorphic and antiholomorphic sectors. In the next
section we show to what extent study of string dynamics  confirms these
conclusions.

\chapter{\bf Amplitudes in String Field Theory}

In the previous section we have seen that although string field theory
 deals with an enormous number of fields on each level, in two dimensions
 the picture simplifies a lot. In fact, only one field degree of
 freedom (tachyon) and certain discrete states remain. It is
 natural to try to confirm this important conclusion by studying the
 scattering amplitudes for the (on-shell) tachyons. Indeed, by unitarity,
 poles of the tree level amplitudes
should correspond to the physical values of momenta. This is the task
 which we now undertake.

Calculations are easiest in Siegel's gauge, and this is the gauge we
 use in this section. However, one has to pay a price for that: discrete
 states here differ from those described in Sec. 4. Although the
 values of physical momenta are still the same, polarization structure
 becomes much more complex. In particular, in Siegel's gauge there
are physical states corresponding to ghost excitations (see below).

Three point on-shell tachyon correlators are normalized to unity
 (in our approach, all amplitudes are bulk). The simplest
 non-trivial calculation is the determination of the four-pont
 functions. We calculate the amplitude as a sum over poles,
 since we are interested in finding out which intermediate states
 contribute to it (the full amplitude in the field-theoretic
 formalism was calculated in [\Kaku]). Let us discuss in more detail
 the $s$-channel contribution. It is given by:

$$
A^{(s)} = \, {g^2 \over 2} \, \sum_{i=1} V_{k_2 i k_1} \, D_i \, V_{k_4 i k_3}
$$

\noindent
where by $k_i$ we have denoted on--shell tachyons. Let us consider
 first the contributions up to the first level:

$$
A^{(s)} = \, ({16 \over 27})^(k_1 + k_2 + {Q \over 2})^2 \, (\,
 {1 \over (k_1 + k_2 + \, {Q \over 2})^2} \, + {(N_{10}^{12})^4 \,
[(k_1 - k_2) \cdot (k_3 - k_4)]^2 \over (k_1 + k_2 + \, {Q \over 2})^2
 \, + 2}  \, + \, \cdots \, )
\eqno\eq
$$

\noindent
It is important to note that $\psi$, $\eta$ intermediate state has
 vanishing residue. This is because the potential contribution is
 proportional to the $N^{12}_{10} +
N^{21}_{10}$, which vanishes. This is, of course, in agreement
 with our discussion in Sec. 3 and 4, where it was stressed that the
 only degree of freedom on the first level is a discrete graviton.

Let us now specialize to the case of interest, namely to $A_{3,1}$. In
 that case we have $k_4 = - \sqrt 2$, $k_1 + k_2 + k_3 = \sqrt 2$,
 and $(k_1 + k_2 + {Q \over 2})^2 = 2 - 2 \sqrt 2 k_3$. Also,
 $(k_1 - k_2) \cdot (k_3 - k_4) = 2 \sqrt 2 \, (k_1 - k_2)$, and
 $(N_{10}^{12})^2 = {4 \over 27}$. Therefore, $(5.2)$ becomes:

$$
A_{+ + + -} = \, {g^2 \over 2} \,  ( \, {1 \over 1 - \sqrt 2 k_3} + \,
 {{1 \over 2} (k_1 - k_2)^2  \over 2 - \sqrt 2 k_3} \, + \cdots \, )
\eqno\eq
$$

\noindent
Notice that poles appear only for certain discrete values of momenta
 (as oppose to the continuum of momenta in $d > 2$ case). Furthermore,
 poles are of the generic form $k = {n \over \sqrt 2}$. Note that the pole
structure of the scattering amplitudes agree with that in Sec. 2 after
the appropriate rescaling is performed. It is important to understand that
although it seems that we have less poles here, it is only because we are
performing the calculations in the Euclidean
instead of the Minkowski space-time. Indeed, in the  Euclidean space-time the
anomalous momentum
conservation  \toanot~ is enforced by a dirac delta function since we,
basically
have:

 $$ {1 \over 2 \pi k } \to \delta(k) $$

for the analytically continued values of $k \to ik$. Therefore, in the
Euclidean
space-time, we do not
get poles for the three-point function, whereas for the $4$ point function we
get a single pole in $k_i$.

\noindent
The equation $(6.2)$ can be compared directly, also, with the singular part of
the well-known conformal result [\Fran]:

$$
A_{+ + + -} = \, {\Gamma(1 - \sqrt 2 k_1) \over \Gamma(\sqrt 2 k_1)} \,
{\Gamma(1 - \sqrt 2 k_2) \over \Gamma(\sqrt 2 k_2)} \,
 {\Gamma(1 - \sqrt 2 k_3) \, \over \Gamma(\sqrt 2 k_3)}
\eqno\eq
$$

\noindent
where, again, the momenta and fields are rescaled compared to \strS~.

\noindent
In exactly the same way as it was done in Sec. 2, we expand that expression in
poles of $k_3$. We obtain:

$$
A \sim \, {1 \over 1 - \sqrt 2 k_3} \, + \, {2 k_1^2  \over 2
 - \sqrt 2 k_3} \, \cdots \,
\eqno\eq
$$

\noindent
which coincides with $(6.4)$ if one notice that the momentum
 conservation in that case requires that $k_1 + k_2 = 0$. Thus, we have
established
 the equivalence between the field--theoretic formalism in
 Siegel's gauge, and the effective Lagrangian and string results up to the
first
level.

Now we are ready to undertake more
elaborate task, namely, to compute the second level contribution. In the
 process we establish which fields, in Siegel's gauge, actually contribute to
the spectrum.
Most of the ground work is provided by the results in Sec. 3. One
 reads--off the propagators from $(4.7)$. To calculate the contributions
involving ghosts we bosonize ghosts [\Samu]. Some of the states, such as
$\bar{b}_{-1} c_{-1} \ket{\Omega} = \, \si_{-1} \ket{+2} \ket{\bar{0}}$,
and $\bar{b}_{-2} c_{-2} \, \ket{\Omega} = \, {1 \over 2} \bar{\si}_{-1}
\ket{\bar{0}} \, (\si_{-2} + \si_{-1}^2) \, \ket{+2}$ (and their
 complex conjugated), do not contribute to the 4-point tachyon amplitude
 because of the term $\si_{-1}$. Indeed, they are sandwiched in the
 following generic way: $\bra{V_{3}} \, \ket{k_i} \si_{-1} \ket{\cdots}
\ket{k_j}$.
Note that $\ket{k_i}$ has the ghost number $\la=1$. Calculation of
 the matrix element boils down to calculating the commutator
 $[N_{10}^{rs} \, \si_{1}^{r} \, \si_{0}^{s}, \, \si_{-1}^{(2)}] = \, {1 \over
2}\, (\,  N^{23}_{10} \la^{(3)} \, + \, N^{21}_{10} \, \la^{(1)}\, ) =
 \, 0$. In the last line we have used the fact that $N^{23}_{10} =
 N^{12}_{10} = - N^{21}_{10}$. Using this, it is easy to see that,
 in fact, there is only $9$ states (out of potentially $17$)) which give rise
to
the nonzero residues of the amplitude on that level. These are states
corresponding to the fields: $A_{\mu \nu \chi \la}$, $B_{\mu \nu \chi}$,
 $C_{\mu \nu \chi}$, $D_{\mu \nu}$, $E_{\mu \nu}$, $F_{\mu \nu}$, $J_{\mu}$,
$K_{\mu}$, and $L$. One concludes that in Siegel's gauge a representation
 of the physical states must involve linear combinations of both matter
 and ghost excitations. The second order contribution to the scattering
amplitude is given by:

$$
A \propto \, {{1 \over 32} \, - {1 \over 8} \, (k_1 - k_2)^2 \, + \,
 {1 \over 8} \, (k_1 - k_2)^4 \over 3 - \sqrt 2 k_3}
\eqno\eq
$$

\noindent
As before, we can easily convince ourselves that $(6.6)$ is consistent
 with the full amplitude $(6.4)$. In fact, since momentum--energy
 conservation leads to: $k_1 + k_2 + k_3 = \, \sqrt 2$, and $k_3 =
 {3 \over  \sqrt2}$ is the value of the pole, we have that $k_1 -
k_2 = 2 k_1 + {1 \over \sqrt 2}$. One can rewrite $(5.6)$ as:

$$
A \propto \, {(\sqrt 2 k_1)^2 (1 + \, \sqrt 2 k_1)^2 \over 2
(3 - \sqrt 2 k_3)}
\eqno\eq
$$

\noindent
Exactly the same expression can be obtained from $(6.4)$ since ${\Gamma(1
-\sqrt
2 k_3) \over \Gamma(\sqrt 2 k_3)} \sim {1 \over 4} \,
{1 \over 3 - \sqrt k_3}$, and:

$$ \eqalign
{
&{\Gamma(1 - \sqrt 2 k_1) \over \Gamma(\sqrt 2 k_1)} \, {\Gamma(1
 - \sqrt 2 k_2) \over \Gamma(\sqrt 2 k_2)} =
\, {\Gamma(2 + \sqrt 2 k_1) \over \Gamma(\sqrt 2 k_1)}
{\Gamma(1 - \sqrt 2 k_1) \over \Gamma(-1 -\sqrt 2 k_1)}
 \, = \, \cr
&= \, (1 + \sqrt 2 k_1)^2 \, (\sqrt 2 k_1)^2 \, , \cr
}
\eqno\eq
$$

\noindent
Poles correspond, again, to a discrete value of momentum (in this case,
 $k = {3 \over \sqrt 2}$. To conclude, we have shown that in Siegel's
 gauge the amplitude receives contributions from discrete states.
 These discrete states have, however, different representation than the
 Lorentz gauge states (Sec. 5). Probably more satisfactory approach would
 be to construct the field theory in Lorentz gauge. In that case poles
 would have to correspond to precisely those well-known discrete states.
 That would be, in that sense, a `minimal' string field theory which
 would, possibly bring us closer to our ultimate goal:
 relating BRST field theory with the simple $2d$ models.

\chapter{\bf Concluding Remarks}

In this paper we have attempted to make a step towards clarifying the
relationship between the string effective Lagrangian and the general BRST
string
theory.
Despite the
limitations of the effective Lagrangian approach due to the
ignorance about the exact form of the tachyon and the higher spin field
interactions, it serves as a useful bridge between the simple string models and
the general BRST approach. Although recently some progress has been made
[\Bel],
general string field approach remains mysteriously complicated even in two
dimensions. Moreover, it is quite difficult to study BRST string field theory
in
different backgrounds, or even to account for nontrivial cosmological constant
$\mu$. In that sense, the effective Lagrangian is very useful since one can
there, at least in
principle, easily expand the action around different backgrounds. That is why
we
hope that by understanding better the relationship between the two approaches,
one can gain some insight on how to reformulate the general BRST string field
theory, and make it more attractive to the public.

There are several interesting questions which one would like to address. For
example, one would like to find more precise relation between the effective
Lagrangian and CSFT
interactions, not just the quadratic parts of the action. Also, one could, in
principle, try to generalize the effective Lagrangian to include higher spin
field interactions.
Although that was argued in the text to be quite tedious, perhaps it could shed
some new light on the general structure of CSFT. As an another interesting
exercise, one may suggest performing of scattering amplitude calculations
similar to Sec. 6 but in Lorentz gauge (see Sec. 5). An advantage of such an
approach would be to directly see which of the well--known DS actually
contribute to scattering. There are some indications that, in fact, only
positively dressed states contribute to $S$ matrix, while negatively dressed do
not. Finally, perhaps the most important problem would be to establish the
precise connection between the above approaches and the collective field
theory.
At present, however, we can say very little about that important issue.

\noindent
{\bf Acknowledgments.} Authors would like to thank A. Jevicki for useful
discussions and his support
throughout the work on the manuscript.

\vfill
\endpage

\refout

\endpage
\end